# ENERGY EFFICIENT SPARK PLASMA SINTERING: BREAKING THE THRESHOLD OF LARGE DIMENSION TOOLING ENERGY CONSUMPTION

Charles Manière[a], Geuntak Lee[a,b], Joanna McKittrick[b], Eugene A. Olevsky[a,c],*,**

(a) Powder Technology Laboratory, San Diego State University, San Diego, USA
(b) Mechanical and Aerospace Engineering, University of California, San Diego, La Jolla, USA
(c) NanoEngineering, University of California, San Diego, La Jolla, USA

**Abstract**
An energy efficient spark plasma sintering method enabling the densification of large size samples assisted by very low electric current levels is developed. In this method, the electric current is concentrated in the graphite foils around the sample. High energy dissipation is then achieved in this area enabling the heating and full densification of large (alumina) parts (Ø 40 mm) at relatively low currents (800 A). The electrothermal mechanical simulation reveals that the electric current needed to heat the large samples is 70 % lower in the energy efficient configuration compared to the traditional configuration. The presence of thermal and densification gradients is also revealed for the larger size samples. Potential solutions for this problem are discussed. The experiments confirm the possibility of full densification (96-99 %) of large alumina samples. This approach allows using small (and low cost) SPS devices (generally limited to 10-15 mm samples) for large size samples (40-50 mm). The developed technique enables also an optimized energy consumption by large scale SPS systems.

___________________________

* Corresponding author: **EO**: *E-mail address*: eolevsky@mail.sdsu.edu
**Fellow of the American Ceramic Society

## 1. Introduction

The spark plasma sintering (SPS) process has demonstrated a high potential for the fabrication of high performance materials ranging from polymers, metals to ultrahigh temperature ceramics [1–5]. This process is capable of rendering high temperatures (2500 °C), high heating rates (1000 K/min) and high pressures (150 MPa) which enable: sintering temperatures about 400 K lower than under free sintering, full sintering with grain size retention and the possibility to adjust sample microstructures [5–8]. This process can also be extended to the fabrication of fully dense complex shapes [9–14] and the flash sintering phenomena [15–20] thanks to the combination of externally applied pressure and high level of electric current (1000-10000 A).

The SPS process' industrial implementation requires addressing these three main issues: i) enabling complex net-shaping, ii) increase of productivity and iii) scaling up of the process. Concerning the fabrication of complex shape components, the reduction of the densification and microstructure inhomogeneities is the key point. The non-homogeneous densification caused by the component thickness differences [10] can be solved by multiple punches [12] approach with the possibility to use sacrificial materials [9] or by the "deformed interface approach" [11] where complex shapes can be produced inside a simple cylindrical tooling. The reduction of the temperature gradients inside the SPS tooling [21–23] and inside the processed samples requires an adjustment of the tooling and of the electric current path [24], which can be accomplished with the help of finite element simulations. A special attention should be paid to the predominant role of the electric and thermal contacts at all the tooling interfaces in the formation of the overall temperature field [25–32]. On the other hand, the productivity and the scale up of the SPS process represent a similar challenge because increasing the productivity requires the simultaneous processing of multiple samples [5] which, in turn, introduce large processed volume dimensions. In both cases a high electric power supply is needed. However, the commercial SPS systems' upper current limit is around 40 kA allowing to sinter samples of about 40 cm diameter[33], - a quite low dimension considering the amount of electric current involved.

To overcome the problem of wasteful energy consumption, the overall SPS process has to use more efficiently the high energy delivered by the power supply. In sinter-forging configuration it is possible to consolidate ultra-high temperature materials at low current and for a very short period of time[15, 16, 34]. In this approach, the electric current is concentrated in a graphite felt rendering the heating of the sample that becomes highly dissipative at high temperatures. This process has been used for processing $ZrB_2$[15], $SiC$[19, 34, 35], $ZrO_2$[16], Nd–Fe–B[36] and for sample sizes up to 60 mm. The main drawback of this method is its sinter-forging configuration which represents a limitation on the fabrication of complex shapes. A different technique introduced by Zapata-Solvas *et al*[17] used a graphite die electrically insulated in order to concentrate the electric current in the sample where a similar flash sintering phenomenon occurs allowing a final cylindrical shape (16 mm) to be obtained under a current ranging from 350 to 1200 A. A similar approach was employed to fully sinter SiC [37]. We recently showed that an electric current concentration in and around the sample enabled a stable and ultra-rapid densification of materials from isolative (like $Al_2O_3$) to conductive (like Ni)[38] systems.

In the present study, an energy efficient method, where the electric current lower than 1000 A is constrained in the graphite foil around the sample, is employed to fully sinter up to 40 mm diameter alumina sample. The feasibility of this method is first tested by an electrothermal-mechanical simulation and then experimentally verified.

## 2. Technical Concept and Experimentation

### 2.1. Traditional vs. energy efficient configurations

In this work, three configurations called "traditional" and "energy efficient" and "spiral" will be considered. The difference between these three configurations is in the electric current path in the area of the punches, sample and die. In the traditional configuration, as reported in Figure 1, the punch/die and sample/die interfaces are separated by a graphite foil inserted in a 0.2 mm gap. In this case, the electric current injected by the top punch can move freely across the punch/die interface to go through the die avoiding the alumina powder (electrically insulative) and can be evacuated by the lower punch. This configuration is widely employed in most of the SPS tests reported in the literature.

In the energy efficient configurations (applied to the sample diameter of 30 and 40 mm, Figure 1), two lateral graphite foils are inserted in a 0.4 mm gap between the punch/die and sample/die interfaces, and the external "graphite foil/die" interface is coated with a boron nitride spray to electrically insulate the die. The electric current is then forced to get across a very narrow interface at the level of the sample/die junction. This small area becomes very dissipative compared to the traditional configuration where the electric current is distributed in the die. This configuration allows the generation of high temperatures for a very low amount of electric currents.

For the 30 mm geometry, a different energy efficient configuration (so called "spiral" configuration), where the electric current is confined both in the sample/punch and sample/die graphite foil, has been investigated too. As reported in Figure 2, this configuration can be tested by placing different electric insulations on the punches' surfaces with BN spray to force the electric current to flow in the sample/punch graphite foil with a spiral shape to enhance the tortuosity of the current path and the current density. In this "spiral configuration", the generated thermal energy completely surrounds the sample, thereby promoting more homogeneous and more efficient heating. These spiral graphite foils are inserted directly on the powder bed and do not require cold pressed specimens.

### 2.2. Experimental procedure

All the experiments were performed using a Spark Plasma Sintering System (SPSS DR.SINTER Fuji Electronics model 515). This SPS device has the maximum current of 1500 A, voltage of 20 V and a maximum applied force of 50 kN. In the traditional configuration, the maximum recommended sample diameter is 15 mm which is adapted to conduct the consolidation of any kind of powder and the study of their material properties; this size limitation, however, constrains the elaboration of real-world large size components. An alumina powder $Al_2O_3$ (Cerac, $Al_2O_3$ 99.99% pure, 37 nm) has been employed to generate 15, 30 and 40 mm diameter pellets. All the utilized graphite tooling uses the industrial grade Isocab I-85. The experimental thermal cycle imposed at the die surface included an initial dwell of 580 °C for the pyrometer detection (Chino, IR-AHS2), a 50 K/min ramp up to 1400 °C and a holding time of few minutes which was interrupted when the displacement curve attained a plateau indicating a nearly complete densification. For the simulations study, the equivalent cycle is reported in Figure 3 assuming a green relative density of 50%. The pressure (40 MPa and 30 MPa for the 30 mm and 40 mm diameter specimens, respectively) was applied from the start and maintained up to the releasing of the current. The fractured specimens were analyzed by scanning electron microscopy (FEI Quanta 450, USA).

## 3. Theory and calculation

An electrothermal-mechanical (ETM) simulation of the SPS process has been employed to study and compare the efficiency of the traditional and energy efficient configurations shown in Figures 1. The ETM simulation encompasses the process Joule Heating[26], the powder compaction of the sample based on the continuum theory of sintering[39] whose main equation is detailed in equation (1), the surface to surface thermal radiation[40] and the proportional integral derivative (PID) regulation[32].

$$\underline{\sigma} = \frac{\sigma_{eq}}{\dot{\varepsilon}_{eq}}\left(\varphi\underline{\dot{\varepsilon}} + \left(\psi - \frac{1}{3}\varphi\right)tr(\underline{\dot{\varepsilon}})\mathbb{I}\right) \qquad (1)$$

where $\underline{\sigma}$ is the stress tensor, $\underline{\dot{\varepsilon}}$ the strain rate tensor, $\sigma_{eq}$ and $\dot{\varepsilon}_{eq}$ respectively the equivalent stress and strain rate, $\mathbb{I}$ the identity tensor, $\varphi$ and $\psi$ respectively the shear and bulk moduli.

The details of the ETM properties for graphite and alumina powder and the boundary conditions can be found in Refs.[10, 26, 32, 41, 42]. The complete set of temperature dependent electric and thermal contact resistances (ECR and TCR) values for all the "SPS column" interfaces has been experimentally determined in Ref[25]. As suggested by the previous experimental study[43], the powder/die friction between the alumina powder and the graphite die wall can be neglected.

## 4. Results and discussion

In this section, the beneficial potential of the energy efficient configuration is first quantified by the ETM simulation; then the homogeneity and the distribution of the main ETM fields are discussed. The energy efficient configurations are tested experimentally, the stability and homogeneity of the resulting densification and microstructures are compared with the simulations results. Finally, the optimization of this method by thermal confinement is investigated by the ETM simulation.

### 4.1. Traditional vs energy efficient mode comparison

The ETM simulation allows the comparison of all the aspects of the Joule heating and the densification of the alumina specimens for diameters of 15 to 40 mm. The calculated shrinkage curves for the three specimens' diameters (Figure 1 configurations) is reported in Figure 3. Figure 4 shows the voltage and current curves of the three SPS geometries reported in Figure 1 in the traditional configuration (15 mm diameter specimen) and in the energy efficient configuration for the largest size configurations (30 and 40 mm diameter specimens). The curves for the traditional configurations indicate close voltage values and large differences in electric currents. For the 15 mm specimen diameter, the values are in the typical "SPS 515" range of the electric current magnitude but for larger diameters (30 and 40 mm) the maximum values of the electric current are between 2000 and 2800 A, which are the values significantly higher than the SPS Dr. Sinter 515 electric current limit of 1500 A. This confirms the validity of the SPS Dr. Sinter 515 recommendation of a typical maximum specimen diameter of 15 mm. To overcome this limit, the energy efficient configuration is employed, and it indicates a small increase of voltage of about 30 % above the average value of the traditional configurations and, more importantly, a drastic 70 % lower current. The "punches, sample, die" volume is 24, 125 and 242 cm$^3$ for the 15, 30 and 40 mm specimen diameter configurations, respectively. The energy efficient configurations implemented for 30 and 40 mm specimens introduce a very large amount of graphite volume compared to the 15 mm case but, thanks to the electric current path optimization, the overall electric current values are of a similar or lower magnitude compared to the traditional 15 mm configuration. The difference of consumed electric power between the traditional and energy efficient configuration (5500 *vs* 3900 W for the 40 mm diameter simulations) is less intense than the difference of the current (2800 *vs* 900 A for the 40 mm diameter simulations). The increase of the voltage due to the complex current path and the electric contact resistances is responsible for the smaller difference in power *vs* current. Experimentally, the impact of the tooling on the voltage needs to be considered[44]. Nevertheless, the improvement of the very high currents required to heat large tooling is the main interesting aspect of the energy efficient configuration for the SPS technology.

The current lines and electric volumetric loss density (Joule heating) reported in Figure 5 for the three configurations shown in Figure 1 provide better understanding of the energy efficiency for large size samples. In the traditional configuration (15 mm specimen on the left, Figure 5a), the current lines are distributed in all the punches, graphite foil and die zones and the main dissipative areas are in the punches where the cross-section area is minimal. For the energy efficient configurations, the current lines in Figure 5b, 5c are constrained in the graphite foil at the level of the sample/die interface. This area becomes then highly dissipative (see zoom in Figures 5b, 5c) and enables the heating of the die, sample and punches area for a very low current regardless of the sample dimensions (very small difference exists between the 30 and 40 mm cases as shown in Figures 4 and 5). However, considering the resulting temperature and the relative density fields reported in Figure 6, this highly dissipative area is responsible of some inhomogeneities in the sample depending on the diameter of the specimen. For the 15 mm specimen, the sample temperature difference is about 60 K; for the 30 and 40 mm energy efficient configurations the sample temperature differences increase to 150 and 425 K, respectively. For the 40 mm diameter configuration, this thermal inhomogeneity is sufficient to start creating the densification difference of about 14 %. These gradients originate from the mode of the heat evacuation in the central part. The main heat generation occurs in the sample/die interface, as shown in Figure 5b. The heat generated has then the tendency to accumulate in the die where a large thermal contact resistance is present at the internal punch/die interface[25], and a graphite felt is used at the external die surface. The maximum simulated temperature in the graphite foil for the energy efficient configuration (30 and 40 mm diameter) is 1477 °C, which is 77 K higher than the die temperature. The heat generated in the graphite foil is quickly diffused in the adjacent die and the powder, which limits the maximum temperature in the foil. In the punches' and sample area, the heat generated propagates inside the sample and is evacuated by the punches. This cooling through the punches is responsible for a thermal gradient where the edge of the sample is at a higher temperature compared to the center which is more efficiently cooled via the punches.

*4.2. Experimental verification*

The experimental validations of the theoretically analyzed energy efficient configurations are presented in this section. All the energy efficient and spiral configurations are shown in Figure 7.

The "spiral configuration" shows a very high efficiency of the heating (1800 W) with electric current values of about 300 A up to 1000 °C. After reaching 1000 °C the nature of the electric contact suddenly deteriorates and the electric current rises to the values close to 1500 A (for an electric consumption 6000 W). However, the densification of the alumina sample up to 93 % was possible. For this configuration, the abrupt amount of current generates conditions for local temperature higher than the melting point of alumina 2072 °C. The 30 and 40 mm energy efficient configurations exhibit close values of electric current at the dwell values of about 800 A similar to those obtained by the simulations (see Figure 4). The consumed electric power is about 3500 W which is not far from the 4000 W simulated. For the 40 mm case, some limited instabilities appear in the proximity of the dwell. The final samples' overall relative densities are 99 % for the 30 mm sample and 96 % for the 40 mm sample. This is consistent with the presence of the thermal and densification gradients in the 40 mm sample as suggested by the simulations reported in Figure 6c.

The fracture surfaces of the three samples are shown in Figure 8. In accordance with the temperature and relative density distributions (Figure 6) modeled for the 30 and 40 mm energy efficient configurations, a microstructure with higher densification and grain size is observed on the edge of both samples. The simulation predicts temperatures differences of 150 and 425 K for the 30 mm and 40 mm samples. The grain size of the 30 mm sample is close to 2 µm in the center and 5 µm on the edge; 1 µm and 2 µm are observed for the 40 mm sample in the center and the edge, respectively. The 30 mm sample is 99 % dense and experienced higher and homogeneous heating leading to stronger sensitivity to grain growth; the 40 mm sample has the high temperature and relative density gradients that postpone the overall sample densification and, in turn, the grain growth that is restrained by the grain-boundary-pinning porosity[45, 46]. In accordance with the simulation results shown in Figure 6c, a small level of porosity remains in the center of the 40 mm sample. Despite the instability of the electric current curve shown in Figure 7, the spiral sample has a similar microstructure in the center and the edge; the only inhomogeneity is the small melted area at the edge that generates a "tree leaves" microstructure which looks like dendritic. This fact indicates that the spiral foil approach partially succeeds in homogenizing the temperatures but this configuration is that much sensible to the properties of the electric contacts, so that it can generate overheating.

*4.3. Optimization of energy distribution by thermal confinement*

In order to improve the microstructures homogeneity for the large samples, a method of thermal confinement can be explored. As shown in Figure 9, the heat generation at the level of the sample/die interface can be confined in the sample, reducing the punches cooling by increasing the sample/punch thermal contact resistance (TCR). Figure 9 indicates that increasing the simulated TCR (very low because of the high applied pressure[25]) 10×, 100× and 1000× times results in the nearly complete homogenization of the sample temperature. This configuration is similar to the energy efficient mode presented above, but a highly thermally isolative material needs to be inserted at the sample/die interface to increase the equivalent TCR to the minimum of 100 × its natural TCR value, and this way to homogenize the temperature. This approach represents some difficulties because most of the highly thermally isolative materials have fibrous or highly porous structures[47] and these structures can be destroyed by the SPS high pressure and temperature. Nevertheless, the described approach lends impetus for further development of technical concepts of temperature homogenization.

## 5. Conclusions

In this work, various new energy efficient spark plasma sintering methods allowing the use of a large size tooling at low electric currents are presented. In the described approaches, the traditional electric current path is modified to concentrate the current in a very narrow area, such as the graphite foil volume near the sample's surface. This area dissipates a high amount of energy and acts like a thermal element that can heat very large volumes of processed materials. Using these approaches, it has been possible to sinter alumina specimens of 30 and 40 mm diameter under applied currents below 800 A (70 % lower than the equivalent value reaching the same level of densification in the traditional configuration). Both the electrothermal-mechanical simulation and experiment reveal that the magnitude of thermal and densification gradients in the specimens increase with the increase of their dimensions. The maximum temperature is located at the edge of the sample in the graphite foil which acts as a heating element. The locally higher temperature at the edge of the sample (radial thermal gradient) has an impact on the densification and the grain growth. Indeed, higher grain sizes are also observed at the edges of the processed samples where the temperature is expected by the simulations to be higher. This phenomenon can be reduced by using the thermal confinement of the heat generated in the sample. The experiment utilizing a spiral-shape foil shows even lower values of the electric currents, however, some instabilities may appear if the electric contact at the level the graphite foil evolves creating hot spot area and local melting of alumina.

Besides the advantages of using a low-power SPS system to fabricate larger samples, the energy efficient configurations make feasible also the manufacture of very large scale samples that are so far limited in size to 40 cm under the maximum electric current level of 40 kA.

**Acknowledgements**

The support of the US Department of Energy, Materials Sciences Division, under Award No. DE-SC0008581 is gratefully acknowledged.

**Figure captions**

Fig. 1: Three spark plasma sintering (SPS) configurations employed for samples with diameters of 15, 30, 40 mm; the 15 mm diameter sample is used in the traditional SPS configuration, for larger sample diameters, a double graphite foil layer is used with a boron nitride coating applied at the external lateral graphite foil interface with the purpose of concentrating the electric current in the sample/die foil which makes the overall configuration very dissipative (energy efficient configuration).

Fig. 2: Spiral configuration; all the internal interfaces are covered with boron nitride spray (electrically isolative), the electric current inlet/outlet is in the punches and the current is concentrated in the graphite foil of a spiral shape adjacent to the sample; the expected electric current path is represented in the middle; on the right: the photograph of the spiral configuration at 800°C.

Fig. 3: (upper) imposed thermal cycle at the die surface and resulting sample shrinkage for the three modeled configurations; (lower) imposed relative density for all simulations.

Fig. 4: Simulated voltage (upper), current (middle) and consumed electric power (lower) in the traditional (left) and energy efficient (right) configurations.

Fig. 5: Simulated current lines and electrical volumetric loss density for a) the traditional (Ø 15 mm) and (b, c) energy efficient (Ø 30; 40 mm) configurations.

Fig. 6: Simulated temperature and relative density field for a) the traditional (Ø 15 mm) and (b, c) energy efficient (Ø 30; 40 mm) configurations.

Fig. 7: Experimental analysis of the "spiral" and "energy efficient" configurations; the temperature, current, consumed electric power, applied force and sample displacement curves are shown.

Fig. 8: SEM of fracture surface for the energy efficient configurations.

Fig. 9: Thermal confinement-based optimization of the sample temperature homogeneity for the Ø 40 mm configuration; the factor indicates how many times the punch/sample thermal contact resistance is multiplied in the respective simulations.